\documentclass[prl,twocolumn]{revtex4}
\usepackage{graphicx}% Include figure files
\usepackage{dcolumn}% Align table columns on decimal point
\usepackage{multirow}% Columns spanning multiple rows
\usepackage{amsmath,amsthm,amssymb}
\usepackage{subfigure}
\usepackage{hyperref}

\def\ep{\varepsilon}
\def\ra{\rightarrow}
\def\Chi{\mathcal{X}}

\def\d{\mathpzc{d}}
\def\a{\mathpzc{A}}
\def\f{\mathpzc{f}}
\newcommand{\inner}[2]{\left( #1 \left| #2 \right. \right)}

\DeclareMathAlphabet{\mathpzc}{OT1}{pzc}{m}{it}

\begin{document}
\title{What drives AdS unstable?}

\author{Maciej Maliborski}
\email{maliborski@th.if.uj.edu.pl}
\affiliation{M. Smoluchowski Institute of Physics, Jagiellonian University, 30-059 Krak\'ow, Poland}

\author{Andrzej Rostworowski}
\email{arostwor@th.if.uj.edu.pl}
\affiliation{M. Smoluchowski Institute of Physics, Jagiellonian University, 30-059 Krak\'ow, Poland}
\date{\today}
\begin{abstract}
  We calculate the spectrum of linear perturbations of standing wave
  solutions discussed in
  \href{http://dx.doi.org/10.1103/PhysRevD.87.123006}{[Phys.~Rev.~D~\textbf{87},~123006~(2013)]},
  as the first step to investigate the stability of globally regular,
  asymptotically AdS, time-periodic solutions discovered in
  \href{http://dx.doi.org/10.1103/PhysRevLett.111.051102}{[Phys.~Rev.~Lett.~\textbf{111}~051102~(2013)]}.  We
  show that while this spectrum is only asymptotically nondispersive
  (as contrasted with the pure AdS case), putting a small standing
  wave solution on the top of AdS solution indeed prevents the
  turbulent instability.  Thus we support the idea advocated in
  previous works that nondispersive character of the spectrum of
  linear perturbations of AdS space is crucial for the conjectured
  turbulent instability.
\end{abstract}

\maketitle
%%%%%%%%%%%%%%%%%%%%%%%%%%%%%%%%%%%%%%%%%%%%%%%%%%%%%%%%%%%%%%%%%%%%%%%%%%%%%%
\textit{Introduction}.
%%%%%%%%%%%%%%%%%%%%%%%%%%%%%%%%%%%%%%%%%%%%%%%%%%%%%%%%%%%%%%%%%%%%%%%%%%%%%%
A recent numerical and analytical study suggest that anti-de Sitter
(AdS) space is unstable against the formation of a black hole under a
large class of arbitrarily small perturbations \cite{br, jrb, dhs,
  bll}. It is argued \cite{piotr_GR20, mr_IJMPA} that the two crucial
ingredients of the mechanism of instability are: (1) the lack of
dissipation of energy by radiation to null infinity (opposite to the
Minkowski case) and (2) a resonant (nondispersive) spectrum of linear
perturbations of AdS. This means that at linear level wave packets do
not disperse in AdS and, once their modes get coupled through
nonlinearities (coming either from self-gravity or self-interaction),
it leaves a long time for them to interact. Then the (conserved)
energy is efficiently transferred into higher and higher frequencies
i.e. gets concentrated on finer and finer spatial scales that
ultimately leads to a black hole formation. On the other hand it was
suggested in \cite{br, dhs, dhms} that there exist asymptotically AdS
(aAdS) solutions that, being arbitrarily close to AdS, are immune to
this instability. Indeed, two explicit examples of such stable
stationary aAdS solutions were given, namely time-periodic solutions
\cite{mr_PRL} and standing waves \cite{bll_BS} (refereed to as boson
stars by the authors) for real and complex massless scalar field
respectively. The existence of this type of solutions seems to be a
rule rather then an exception. This suggests that while the AdS space
itself is unstable against black hole formation, putting some
fine-tuned small ripples on AdS can prevent the instability
\cite{dhms, mr@bll}. But what makes these ripples stable?  In this
note we support the conclusions of the work \cite{dhms} that a
nondispersive character of the spectrum of linear perturbations on the
fixed AdS background is crucial for the assumed instability.
% In this note we argue that a nondispersive character of the spectrum
% of linear perturbations on the fixed AdS background may be crucial
% for the assumed instability.
Firstly, up to now, there was some clash between analytical
\cite{dhms} and numerical results of one of us \cite{m} to what extent
only asymptotically resonant character of the spectrum is good enough
to trigger the instability. We refined the numerical analysis of
\cite{m} and found that asymptotically resonant spectrum is not
sufficient to trigger instability for small perturbations.  Secondly,
to investigate what makes the time-periodic solutions discovered in
\cite{mr_PRL} stable we start with a simpler case of standing waves
discussed in \cite{bll_BS}.  Namely, we investigate in detail the
spectrum of their linear perturbations.  We show that, while this
spectrum is only asymptotically resonant, putting a small standing
wave in the center of AdS prevents the instability.

%%%%%%%%%%%%%%%%%%%%%%%%%%%%%%%%%%%%%%%%%%%%%%%%%%%%%%%%%%%%%%%%%%%%%%%%%%%%%%
\textit{Flat space enclosed in a cavity, revisited}.
%%%%%%%%%%%%%%%%%%%%%%%%%%%%%%%%%%%%%%%%%%%%%%%%%%%%%%%%%%%%%%%%%%%%%%%%%%%%%%
In \cite{m} a spherically symmetric self-gravitating massless scalar
field enclosed in a perfectly reflecting spherical cavity was studied
as a toy model for the assumed AdS instability. This somewhat
artificial model allowed for two types of reflecting boundary
conditions: Dirichlet and Neumann, resulting in strictly resonant
(nondispersive) spectrum $\omega_j=j\pi/R$ and only asymptotically
resonant spectrum $\tan R\omega_j = R\omega_j$, respectively (here $R$
stands for the cavity radius). The resonant case, being a close
analogue of the AdS case, showed a perfect scaling with the amplitude
of the initial perturbation (compare the Fig.~2 in \cite{m}, with the
key numerical evidence for AdS instability, the Fig.~2 in \cite{br})
and the similar behavior of energy spectra to the AdS case (compare
the Fig.~4 in \cite{m} with the Fig.~2 in \cite{mr_IJMPA}) and
strengthened the evidence for a robust mechanism of instability
sketched in \cite{br}. In spite of the fact that the analogous scaling
in the Neumann boundary case (cf. Fig.~5 in \cite{m}) might have not
seem compelling enough, it was concluded in quest of further
robustness that \textit{...~the spectrum of linearized perturbations
  need not be fully resonant for triggering the instability}.
\begin{figure}[h]
  \centering
  \includegraphics[width=1.03\columnwidth]{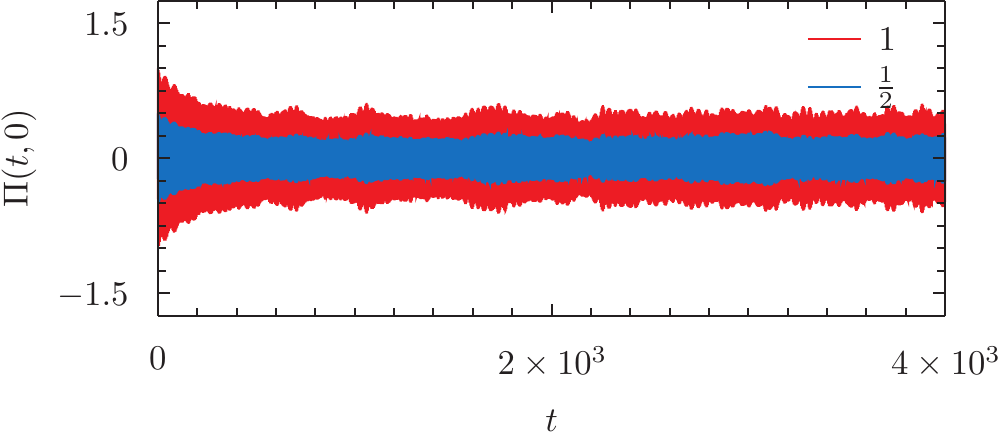}
  \\[2.5ex]
  \includegraphics[width=0.99\columnwidth]{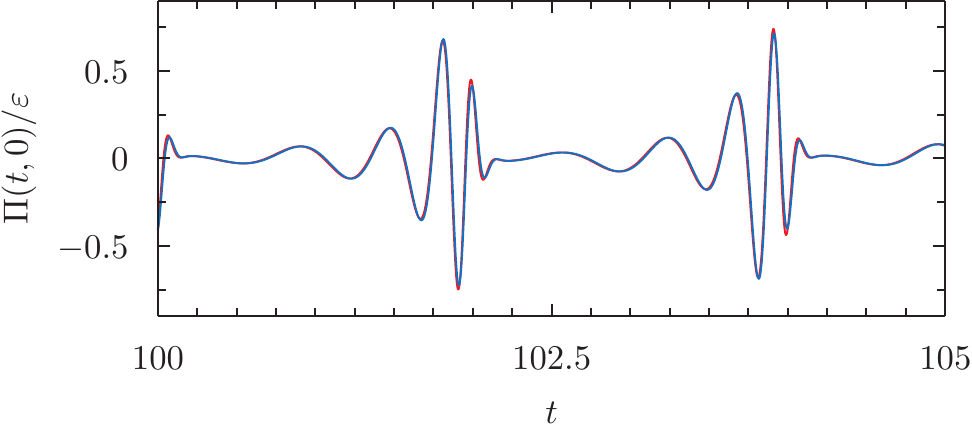}
  \\[2.5ex]
  \includegraphics[width=0.99\columnwidth]{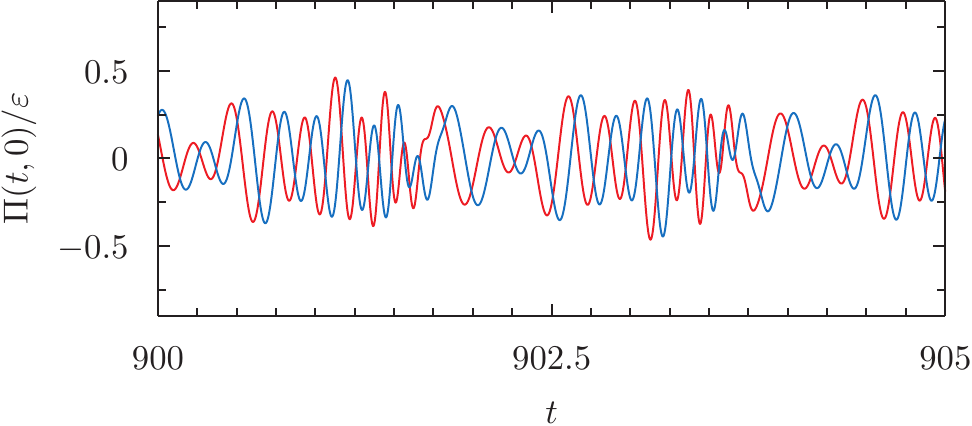}
  \caption{\textit{Top}. The function $\Pi(t,0)$ for solutions to the
    model \cite{m} with Neumann boundary condition at the cavity
    boundary (of size $R=1$) with initial data
    (\ref{eq:box_initial_data}) for small amplitude shows very
    different behaviour as opposed to moderate and large
    perturbations, see Fig.~5 in \cite{m}. The spectral code
    \cite{m_phd} conserves the total mass up to $2.5\times 10^{-15}$
    over long integration times.  \textit{Middle}. The closeup showing
    scaling of $\Pi(t,0)$ function with an amplitude of the
    perturbation $\ep$ (with the same color
    coding).  \textit{Bottom}. Due to the dispersive spectra for
    Neumann boundary condition the initially localized preturbation
    spreads over the entire spatial domain which prevents the
    collapse. For late time there is also phase shift between the
    signals of different amplitudes.}
  \label{fig:ricci_minkowski}
\end{figure}
On the other hand the authors of \cite{dhms} came to the opposite
conclusion based on nonlinear perturbation analysis. The clash between
those two statements became even more prominent with the discovery of
concrete examples of (nonlinearly) stable aAdS solutions
\cite{mr_PRL,bll_BS}, previously advocated in \cite{dhms} and the
question what makes them immune to the instability discovered in
\cite{br}. Thus we ran the simulation for the same family of initial
data as \cite{m}
\begin{equation}
  \label{eq:box_initial_data}
  \Phi(0,r) = 0\,, \qquad
  \Pi(0,r) = \ep\exp\left(-64\tan^{2}\frac{\pi}{2}r\right)\,,
\end{equation}
but still smaller amplitudes \footnote{there is a typo in the width of
  the gaussian in the eq. (13) of \cite{m}: the coefficient in the
  exponent should read 64 instead of 32}. The results are depicted in
Fig.~1. For the Dirichlet boundary condition we confirmed the scaling
depicted in the Fig.~2 of \cite{m}: the scaling works better when the
amplitude of the initial data is decreased. For the Neumann boundary
condition we found that scaling does not improve as we decrease the
amplitude, while for $\ep \lesssim 1$ the instability is not triggered
at all.

%%%%%%%%%%%%%%%%%%%%%%%%%%%%%%%%%%%%%%%%%%%%%%%%%%%%%%%%%%%%%%%%%%%%%%%%%%%%%%
\textit{Standing waves in AdS}.
%%%%%%%%%%%%%%%%%%%%%%%%%%%%%%%%%%%%%%%%%%%%%%%%%%%%%%%%%%%%%%%%%%%%%%%%%%%%%%
In this section we revisit the problem of nonlinear stability of
standing waves (boson stars) in AdS and show that while the numerical
results of \cite{bll_BS} provide the evidence for their stability, the
spectrum of their linear perturbation is only asymptotically
resonant. To make this note self-contained we rewrite the equations
for a complex, selfgravitating massless scalar field with negative
cosmological constant studied extensively in \cite{bll,bll_BS}. We
parametrize the $(d+1)$--dimensional asymptotically AdS metric by the
ansatz
\begin{equation}
  \label{eq:adsd+1_ansatz}
  ds^2\! =\! \frac {\ell^2}{\cos^2{\!x}}\left( -A e^{-2 \delta} dt^2 +
    A^{-1} dx^2 + \sin^2{\!x} \, d\Omega^2_{d-1}\right),
\end{equation}
where $\ell^2=-d(d-1)/(2\Lambda)$, $d\Omega^2_{d-1}$ is the round
metric on $S^{d-1}$, $-\infty<t<\infty$, $0\leq x<\pi/2$, and $A$,
$\delta$ are functions of $(t,x)$. The evolution of the system is
governed by Einstein equations
\begin{align}
  \label{eq:einstein_complex_scalar}
  G_{\alpha\beta} + \Lambda g_{\alpha\beta} & = 8 \pi G
  \left(\partial_{\alpha}\phi\,\partial_{\beta}\bar{\phi} -
    \frac{1}{2}g_{\alpha\beta}\,\partial^\mu\phi\,\partial_\mu
    \bar{\phi}\right),
  \\
  g^{\alpha\beta}\nabla_{\alpha}\nabla_{\beta}\phi & = 0\,,
\end{align}
with $\phi$ standing for the massless complex scalar field.  For the
metric ansatz (\ref{eq:adsd+1_ansatz}) this system boils down to
(using the units with $8\pi G=d-1$)
\begin{equation}
  \label{eq:wave}
  \dot\Phi = \left( A e^{-\delta} \Pi \right)', \quad \dot \Pi =
  \frac{1}{\tan^{d-1}{\!x}}\left(\tan^{d-1}{\!x} \,A e^{-\delta} \Phi
  \right)',
\end{equation}
\begin{align}
  \label{eq:slicing_cnd}
  \delta' \!&=\! - \sin{x}\cos{x}\left( \left|\Phi\right|^2 +
    \left|\Pi\right|^2 \right)\,, \\
  \label{eq:hamiltonian_cnstr}
  A' \!&= \!\frac{d-2+2\sin^2{\!x}}{\sin{x}\cos{x}}\,(1-A) +
  A\delta'\,,
\end{align}
where ${}^{\cdot}=\partial_t$, ${}'=\partial_x$, and
\begin{equation}
  \label{eq:Phi_Pi_definitions}
  \Phi= \phi', \qquad \Pi= A^{-1} e^{\delta} \dot \phi \,.
\end{equation}
Note that the set of equations
(\ref{eq:wave}-\ref{eq:hamiltonian_cnstr}) has the same form as in
\cite{jrb, mr_IJMPA, mr_PRL} with the only exception that auxiliary
fields (\ref{eq:Phi_Pi_definitions}) are now complex valued functions,
and it differs from one presented in \cite{bll} by the scaling factor
$\cos^{d-1}{x}$.  As discussed in \cite{mr_IJMPA} we supply this
system with reflecting boundary conditions
\begin{equation}
  \label{eq:boundary}
  \Pi(t,\pi/2) = 0, \quad \delta'(t,\pi/2) = 0, \quad A(t,\pi/2) = 1,
\end{equation}
to require smooth evolution with a conserved total mass
\cite{mr_IJMPA}.  With the stationarity ansatz
\begin{equation}
  \label{eq:sw_ansatz}
  \phi(t,x) = e^{i \Omega t}\f(x), \ \
  \delta(t,x)=\d(x), \
  A(t,x)=\a(x),
\end{equation}
$\Omega>0$, the system (\ref{eq:wave}-\ref{eq:Phi_Pi_definitions}) is
reduced to
\begin{align}
  \label{eq:sw_f}
  -\Omega^{2}\frac{e^{\d}}{\a} \f & =
  \frac{1}{\tan^{d-1}x}\left(\tan^{d-1}x \a e^{-\d}\f'\right)', \\
  \label{eq:sw_d}
  \d' & = - \sin x\cos x \left[\f'^{2} + \left(\frac{\Omega
        e^{\d}}{\a}\f\right)^{2}\right], \\
  \label{eq:sw_a}
  \a' & = \frac{d - 2 + 2\sin^{2}x}{\sin x\cos x}(1-\a) + \a\d'.
\end{align}
We refer to solutions of the form (\ref{eq:sw_ansatz}) as standing
wave solutions rather than boson star.  For a review on a different
models of boson star solutions and their possible astrophysical and
cosmological relevance see \cite{lp_lrr} and references therein.  We
construct the solutions of the system (\ref{eq:sw_f}-\ref{eq:sw_a})
both numerically and perturbativelly using a modified versions of the
codes for time-periodic solutions \cite{mr_PRL}.  Using perturbative
approach we seek for solution in a form
\begin{align}
  \label{perturbative_series}
  \f(x) \!&= \sum_{\mbox{{\small odd }}\lambda \geq
    1}\ep^{\lambda}\,\f_{\lambda}(x), \quad \f_{1}(x) =
  \frac{e_{\gamma}(x)}{e_{\gamma}(0)}
  \\
  \label{pexp_delta_A}
  \d(x) \!&=\sum_{\mbox{{\small even }} \lambda \geq
    2}\ep^{\lambda}\,\d_{\lambda}(x), \quad 1 - \a(x) =
  \sum_{\mbox{{\small even }} \lambda \geq
    2}\ep^{\lambda}\,\a_{\lambda}(x),
  \\
  \label{Omega}
  \Omega \!& = \omega_{\gamma} + \sum_{\mbox{{\small even }} \lambda
    \geq 2}\ep^{\lambda}\,\omega_{\gamma,\lambda},
\end{align}
where $e_{\gamma}(x)$ is a dominant mode in the solution in the limit
$\ep \ra 0$ ($e_{j}(x)$ is an eigenbasis of a linear problem for a
fixed AdS$_{d+1}$ background $L e_{j}(x) = \omega_{j}^{2}\,e_{j}(x)$,
$L=-\tan^{1-d}{x}\,\partial_{x}\left(\tan^{d-1}{x}\,\partial_{x}\right)$
with eigenfrequencies $\omega_{j}^{2} = (d+2j)^{2}$, $j=0, 1, \ldots$;
for explicit form of $e_{j}(x)$ see e.g. \cite{mr_PRL}). Since the
$e_{\gamma}(x)$ function has exactly $\gamma$ nodes we refer to the
solution with dominant mode $\gamma=0$ as a ground state solution
while for solutions with $\gamma>0$ as excited states (as in the boson
star nomenclature).  This particular choice of $\f_{1}(x)$ together
with a requirement $\f_{\lambda}(0)=0$ for $\lambda\geq 3$ fixes a
value of scalar field at the origin to $\f(0)=\ep$.  Next, at each
perturbative order $\lambda$ we decompose scalar field
$\f_{\lambda}(x)$ and metric functions $\a_{\lambda}(x)$,
$\d_{\lambda}(x)$ in the eigenbasis $e_{j}(x)$ as for the
time-periodic solutions, such that boundary conditions
(\ref{eq:boundary}) are satisfied at each order.  This allow us to
obtain the solution by solving the linear algebraic system for Fourier
coefficients, where the frequency corrections
$\omega_{\gamma,\lambda}$ are fixed by an integrability
conditions.  In this way we get a unique solution, up to arbitrarily
high order $\lambda$, with $d$ and $\gamma$ being the only
parameters. For more details see \cite{mr_PRL}.

\begin{figure}[t]
  \centering
  \includegraphics[width=0.99\columnwidth]{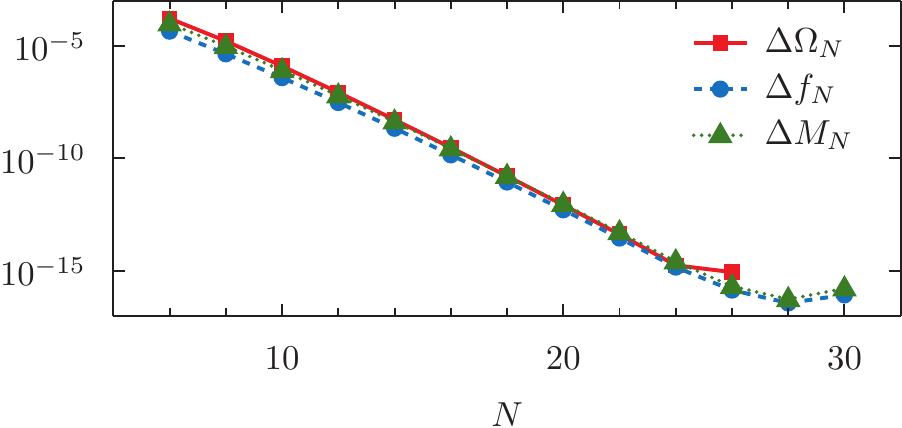}
  \\[2ex]
  \includegraphics[width=0.99\columnwidth]{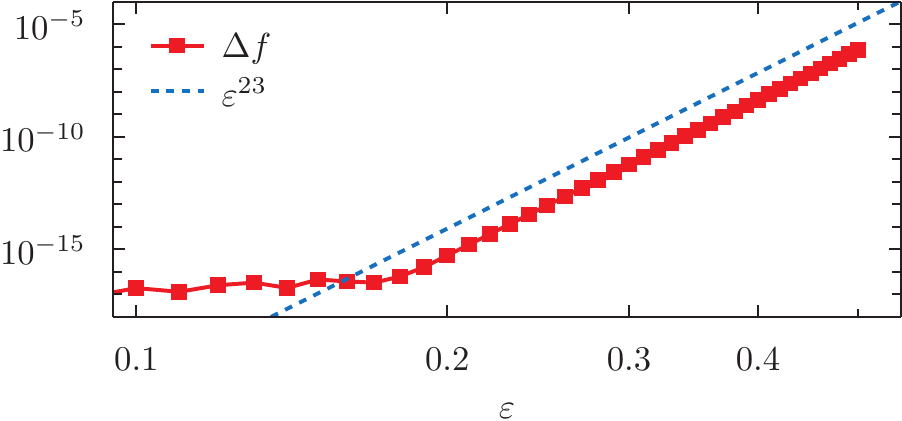}
  \caption{\textit{Top.} The convergence test of numerical code for
    ground state standing wave solution with $\f(0)=3/10$
    ($\mbox{$\Omega\approx 4.56690$}$).  The frequency error
    $\Delta\Omega_{N} = |\Omega_{N}-\Omega_{N=32}|$, scalar field
    profile error $\Delta\f_{N}=||\f_{N}-\f_{N=32}||_{2}$, and total
    mass error $\Delta M_{N}=|M_{N}-M_{N=32}|$ computed for increasing
    number of Fourier coefficients $N$ in
    (\ref{eq:sw_num_f},~\ref{eq:sw_num_da}) compared with reference
    solution with $N=32$. \textit{Bottom.}  The comparison of
    numerical and analytical ground state standing wave solutions for
    varying value of $\f(0)=\ep$. The scalar field absolute error
    $\Delta\f=||\f_{\mathrm{num}} - \f_{\mathrm{pert}}||_{2}$ is
    computed for numerical solution with $N=48$ eigenmodes, the
    perturbative series was found up to
    $\mathcal{O}\left(\ep^{23}\right)$ order.  For small values of
    $\ep<0.2$ the rounding errors dominate.  The discrete $l^{2}$-norm
    $||\,.\,||_{2}$ was computed on as set of equally spaced grid
    points $x_{i}=i\pi/800$, $i=1,\ldots, 400$.}
  \label{fig:test_num_pert}
\end{figure}

As for the time-periodic solutions we also construct standing wave
solution by solving (\ref{eq:sw_f}-\ref{eq:sw_a}) numerically.  Here
we represent the solution by a set of $3N$ Fourier coefficients
\begin{align}
  \label{eq:sw_num_f}
  \f(x) \!&= \sum_{j=0}^{N-1}\hat{\f}_{j} e_j(x),
  \quad \d(x) = \sum_{j=0}^{N-1}\hat{\d}_{j}\left(e_{j}(x) - e_{j}(0)\right), \\
  \label{eq:sw_num_da}
  \a(x) \!& = 1 - \sum_{j=0}^{N-1} \hat{\a}_{j} e_j(x),
\end{align}
and the frequency $\Omega$.  With an approximation (\ref{eq:sw_num_f},
\ref{eq:sw_num_da}) satisfying boundary conditions (\ref{eq:boundary})
we require for the equations (\ref{eq:sw_a}-\ref{eq:sw_f}) to be
satisfied at the set of $N$ collocation points.  Additionally we add
to this system a condition fixing the value of scalar field at the
origin $\f(0)=\ep$.  In this way we get a nonlinear eigenvalue system
of $3N+1$ equations for the same number of unknowns approximating a
solution to (\ref{eq:sw_f}-\ref{eq:sw_a}). Fig.~\ref{fig:test_num_pert}
shows both the convergence rate of our numerical method and
comparison with perturbativelly constructed solution.

%%%%%%%%%%%%%%%%%%%%%%%%%%%%%%%%%%%%%%%%%%%%%%%%%%%%%%%%%%%%%%%%%%%%%%%%%%%%%%
\textit{Linear stability of standing waves}.
%%%%%%%%%%%%%%%%%%%%%%%%%%%%%%%%%%%%%%%%%%%%%%%%%%%%%%%%%%%%%%%%%%%%%%%%%%%%%%
To study the linear stability we make the perturbative ansatz
$(|\mu|\ll 1)$
\begin{align}
 \label{eq:sw_pert_1}
 \phi(t,x) & = e^{i\Omega t}\left(\f(x) + \mu\,\psi(t,x) +
   \mathcal{O}\left(\mu^{2}\right) \right),\\
 \delta(t,x) & = \d(x) + \mu\left(\alpha(t,x) - \beta(t,x)\right) +
 \mathcal{O}\left(\mu^{2}\right), \\
 A(t,x) & = \a(x) \left( 1 + \mu\,\alpha(t,x) +
   \mathcal{O}\left(\mu^{2}\right) \right), \\
\end{align}
and we neglect higher order terms in $\mu$.  Next, we assume harmonic
time dependence of the perturbation
\begin{align}
  \psi(t,x) & = \psi_{+}(x)e^{i\Chi t} + \psi_{-}(x)e^{-i\Chi t}, \\
  \alpha(t,x) & = \alpha(x)\cos\Chi t, \\
  \label{eq:sw_pert_2}
  \beta(t,x) & = \beta(x)\cos\Chi t,
\end{align}
where $\psi_{+}(x)$ and $\psi_{-}(x)$ are both real functions.  This
is the most general ansatz allowing for separation of $t$ and $x$
dependence, making at the same same time the resulting system of
equations relatively simple (cf. \cite{bll_BS}).  Plugging the
(\ref{eq:sw_pert_1}-\ref{eq:sw_pert_2}) into
(\ref{eq:wave}-\ref{eq:Phi_Pi_definitions}) and linearizing about
$\mu=0$ we obtain a set of differential-algebraic equations
\begin{align}
  \label{eq:alpha}
  \begin{split}
    \alpha & = -\sin{2x}\left\{\frac{\Omega}{\Chi}\f
      \left(\psi_{+}'-\psi_{-}'\right) \right. \\
    & \left. \quad + \f' \left[\left(1-\frac{\Omega
          }{\Chi}\right)\psi_{+} + \left(1 +
          \frac{\Omega}{\Chi}\right)\psi_{-}\right]\right\}\,,
  \end{split}
  \\
  \label{eq:beta}
  \beta' & = - \frac{d - 1 - \cos 2x}{\sin{x}\cos{x}}\frac{\alpha}{\a}\,,
  \\
  \label{eq:psi_pm}
  \begin{split}
    \psi_{\pm}'' & = - \frac{d - 1 - \cos{2x} (1 - \a)}{\a\sin{x}\cos{x}}\psi_{\pm}' \\
    & \quad - \left(1 \mp \frac{\Chi}{\Omega}\right)^2
    \left(\frac{\Omega e^{\d}}{\a}\right)^2\psi_{\pm} \\
    & \quad - \frac{1}{2}\beta'\f' + \left(1 \mp \frac{\Chi}{2
        \Omega}\right) \left(\frac{\Omega e^{\d}}{\a}\right)^2\beta\f\,.
  \end{split}
\end{align}
This system supplied with the boundary conditions (inherited from
(\ref{eq:boundary}))
\begin{equation}
  \psi_{\pm}(\pi/2) = 0, \quad \alpha(\pi/2) = 0, \quad \beta'(\pi/2) = 0,
\end{equation}
and the regularity conditions at $x=0$ is a linear eigenvalue problem
with $\Chi$ as an eigenvalue.  In principle, knowing standing wave
solution $\f(x)$, $\a(x)$, $\d(x)$, we could integrate
(\ref{eq:alpha}-\ref{eq:psi_pm}) to obtain a solution in a closed
form.  Since this is not the case here, we again resort on
perturbative method. We expand the unknown functions $\alpha(x)$,
$\beta(x)$, $\psi_{\pm}(x)$ and frequency $\Chi$ in small parameter
$\ep$ (the same as for the standing wave solution
(\ref{perturbative_series}-\ref{Omega}))
\begin{align}
  \label{eq:Chi_pexp_Psi_pexp}
  \Chi & = \sum_{\mbox{{\small even }}\lambda\geq
    0}\ep^{\lambda}\chi_{\lambda}, \quad \psi_{\pm}(x) =
  \sum_{\mbox{{\small even }}\lambda\geq
    0}\ep^{\lambda}\psi_{\pm,\lambda}(x),
  \\
  \label{eq:ab_pexp}
  \alpha(x) & = \sum_{\mbox{{\small odd }}\lambda\geq
    1}\ep^{\lambda}\alpha_{\lambda}(x), \quad \beta(x) =
  \sum_{\mbox{{\small odd }}\lambda\geq
    1}\ep^{\lambda}\beta_{\lambda}(x).
\end{align}
Plugging (\ref{perturbative_series}-\ref{Omega}) and
(\ref{eq:Chi_pexp_Psi_pexp}-\ref{eq:ab_pexp}) into
(\ref{eq:alpha}-\ref{eq:psi_pm}) we demand that the equations are
satisfied at each order of $\ep$.  Moreover, as for the standing wave
solution we expand the unknown functions in eigenbasis $e_{j}(x)$
\begin{align}
  \label{eq:Psi_pexp}
  \psi_{\pm,\lambda}(x) & = \sum_{j\geq 0}\inner{e_{j}}{\psi_{\pm,\lambda}}
  e_{j}(x), \\
  \alpha_{\lambda}(x) & = \sum_{j\geq 0}\hat{\alpha}_{\lambda,j}e_{j}(x),
  \\
  \beta_{\lambda}(x) & = \sum_{j\geq 0}\hat{\beta}_{\lambda,j}
  \left(e_{j}(x)-e_{j}(0)\right),
\end{align}
At the lowest order $\mathcal{O}(\ep^{0})$ the constraints
(\ref{eq:alpha}) and (\ref{eq:beta}) are identically satisfied, while
from (\ref{eq:psi_pm}) we get two linear second order equations
\begin{equation}
  \label{eq:pert_psi_0}
  L\psi_{\pm,0} - (\chi_{0} \mp \omega_{\gamma})^{2}\psi_{\pm,0} = 0\,,
\end{equation}
Using decomposition of $\psi_{\pm,0}(x)$ and orthogonality of the
basis functions
$\inner{e_{i}}{e_{j}}:=\int_{0}^{\pi/2}e_{i}(x)e_{j}(x)\tan^{d-1}{x}\,dx
= \delta_{i,j}$ we get the condition for the frequency $\chi_{0}$
\begin{equation}
  \left\{
    \begin{aligned}
      \omega_{j}^{2} - (\chi_{0} - \omega_{\gamma})^{2}  & = 0, \\
      \omega_{k}^{2} - (\chi_{0} + \omega_{\gamma})^{2}  & = 0.
    \end{aligned}
  \right.
\end{equation}
This system is satisfied when: $\psi_{-,0}\equiv 0$, $\psi_{+,0} =
e_{j}(x)$, and $\chi_{0}=\omega_{\gamma}\pm\omega_{j}$ or
$\psi_{+,0}\equiv 0$, $\psi_{-,0} = e_{k}(x)$, and
$\chi_{0}=-\omega_{\gamma}\pm\omega_{k}$ (there is also the case when
neither of $\psi_{\pm,0}(x)$ is zero, i.e. $\psi_{+,0}(x) = e_{j}(x)$,
$\psi_{-,0}(x) = e_{k}(x)$ with $k$, $j$ such that $d+2\gamma = |k-j|$
holds, but construction of solutions for this choice breaks down at
higher orders, thus we exclude this case).  Taking into account the
form of the ansatz (\ref{eq:sw_pert_2}), due to the $t\ra -t$
symmetry, these two seemingly different cases are in fact
equivalent.  Therefore, it suffices to consider the former case, so as
a solution of the linear system (\ref{eq:pert_psi_0}) we take
\begin{equation}
  \label{eq:pert_psi_0_sol}
  \psi_{+,0}(x) = e_{\zeta}(x),
  \quad \psi_{-,0}(x) = 0,
  \quad \chi_{0}^{\pm} = \omega_{\gamma} \pm \omega_{\zeta}\,.
\end{equation}

Thus, at the lowest order in $\ep$, solution (\ref{eq:pert_psi_0_sol})
specifies a standing wave with $\gamma$ nodes perturbed by a single
eigenmode with $\zeta$ nodes.  Next, at each odd order $\lambda$ the
constraints are solved as follows. The coefficients
$\hat{\alpha}_{\lambda,j}$ are simply given in terms of the
decomposition of the order $\lambda$ of the right hand side of the
equation (\ref{eq:alpha}).  Next, we rearrange the (\ref{eq:beta}) at
the order $\lambda$ to obtain the linear system for the expansion
coefficients of the $\beta_{\lambda}(x)$ function.  For any even
$\lambda$ the system (\ref{eq:alpha}-\ref{eq:psi_pm}) reduces to two
inhomogeneous equations
\begin{equation}
  \label{eq:pert_psi_lambda}
  L\psi_{\pm,\lambda} - (\chi_{0} \mp \omega_{\gamma})^{2}
  \psi_{\pm,\lambda} = S_{\pm,\lambda}\,,
\end{equation}
with source terms $S_{\pm,\lambda}$ depending on the lower order
expansion coefficients in (\ref{perturbative_series}-\ref{Omega}) and
(\ref{eq:Chi_pexp_Psi_pexp}-\ref{eq:ab_pexp}).  Using the
$\psi_{+,\lambda}(x)$ expansion formula (\ref{eq:Psi_pexp}) and
projecting the first equation in (\ref{eq:pert_psi_lambda}) onto the
$e_{i}(x)$ mode we have
\begin{equation}
  \inner{e_{i}}{\psi_{+,\lambda}}
  = \frac{\inner{e_{i}}{S_{+,\lambda}}}{\omega_{i}^{2} - \omega_{\zeta}^{2}},
   \quad i\neq\zeta\,,
\end{equation}
where we have used the definition of $\chi_{0}^{\pm}$ given in
(\ref{eq:pert_psi_0_sol}). For $i=\zeta$ the necessary condition
\begin{equation}
  \inner{e_{\zeta}}{S_{+,\lambda}} = 0\,,
\end{equation}
is satisfied by an appropriate choice of the parameter
$\chi_{\lambda}$, while the free coefficient
$\inner{e_{\zeta}}{\psi_{+,\lambda}}$ is fixed as follows.  We set the
value of $\psi_{+}(x)$ at the origin to unity (we use the fact that
governing equations are linear and we set
$\psi_{+,0}(x)=e_{\zeta}(x)/e_{\zeta}(0)$), then since
$\psi_{+,0}(0)=1$ we require that $\psi_{+,\lambda}(0)=0$ for
$\lambda\geq 2$ which corresponds to taking
\begin{equation}
  \inner{e_{\zeta}}{\psi_{+,\lambda}} =
  - \sum_{i\neq\zeta}\inner{e_{i}}{\psi_{+,\lambda}}e_{i}(0)\,.
\end{equation}
For a second equation in (\ref{eq:pert_psi_lambda}) after projection
on $e_{k}(x)$ mode, we get
\begin{equation}
  \label{eq:psi_2_lambda_ek}
  \inner{e_{k}}{\psi_{-,\lambda}}
  = \frac{\inner{e_{k}}{S_{-,\lambda}}}{\omega_{k}^{2}
    - \left(2\omega_{\gamma}\pm\omega_{\zeta}\right)^{2}},
  \quad k\neq k_{*},
\end{equation}
where $\omega_{k_{*}} = |2\omega_{\gamma}\pm\omega_{\zeta}|$ and the
sign depends on the particular choice of
$\chi_{0}=\chi_{0}^{\pm}$. For
$\chi_{0}=\chi_{0}^{+}=\omega_{\gamma}+\omega_{\zeta}$ the
$k_{*}=d+2\gamma+\zeta>0$ and the condition
\begin{equation}
  \inner{e_{k_{*}}}{S_{-,\lambda}} = 0,
\end{equation}
can always be satisfied by an appropriate choice of a constant
$\inner{e_{k_{*}}}{\psi_{-,\lambda-2}}$ (it is remarkable that at the
lowest nontrivial order $\lambda=2$ the coefficient
$\inner{e_{k_{*}}}{S_{-,\lambda=2}}$ is always zero for any
combination of $\gamma$ and $\zeta$, so we can continue our
construction to arbitrary high order $\lambda$, having exactly one
undetermined constant after solving order $\lambda$, which will be
fixed at higher order $\lambda+2$).  On the other hand for
$\chi_{0}=\chi_{0}^{-}=\omega_{\gamma}-\omega_{\zeta}$ we have $k_{*}
= \frac{1}{2}\left(\left|d + 2(2\gamma - \zeta)\right| - d\right)$
which can be either positive or negative.  For $k_{*}<0$ there are
always solutions to (\ref{eq:psi_2_lambda_ek}) since the denominator,
on right hand side, is always different from zero for any $k\geq 0$,
and the coefficient $\inner{e_{k}}{\psi_{-,\lambda}}$ will be
determined by the formula (\ref{eq:psi_2_lambda_ek}).  The $k_{*}\geq
0$ case is more involved since there are two possibilities: either $d
+ 2(2\gamma - \zeta)\geq 0$ which gives $k_{*}=2\gamma - \zeta\geq 0$
and there are no solutions to (\ref{eq:psi_2_lambda_ek}) since it
turns out that the coefficient $\inner{e_{k_{*}}}{S_{2,\lambda=2}}$ is
nonzero, which leads to contradiction, either $d + 2(2\gamma -
\zeta)<0$ and for $k_{*} = \zeta - 2\gamma - d\geq 0$ the coefficient
$\inner{e_{k_{*}}}{S_{-,\lambda=2}}$ is zero and the unknown
$\inner{e_{k_{*}}}{\psi_{-,\lambda=2}}$ will be fixed at higher order
$\lambda=4$ and we proceed just like for the $\chi_{0}=\chi_{0}^{+}$
case.  To sum up, for $\chi_{0}=\chi_{0}^{+}$ there are solutions for
any choice of $\gamma$ and $\zeta$, while for the
$\chi_{0}=\chi_{0}^{-}$ there exists solutions only for $\zeta >
2\gamma$.

In this way we construct a solution describing a standing wave with
dominant eigenmode $e_{\gamma}(x)$ peturbed (at the linear level) by a
dominant eigenmode $e_{\zeta}(x)$.  Note the (general) ansatz
(\ref{eq:sw_pert_1}-\ref{eq:sw_pert_2}) allows us to perturb a fixed
standing wave with any eigenmode, as opposed to the analysis presented
in \cite{bll_BS}.  The ansatz proposed in \cite{bll_BS} restricts the
form of perturbation, such that it allows for a $\gamma$-node standing
wave to be perturbed by a solution with $\gamma$-nodes only.  For that
reason it is not suitable to find the full spectrum of linear
perturbations.

Solving the higher orders of perturbative equations (in terms of $\ep$
expansion) we get successive approximation to the solution of the
system (\ref{eq:alpha}-\ref{eq:psi_pm}) and in particular for the
eigenfrequences $\Chi$.  Repeating this procedure for successive
values of $\zeta$ we can compute the spectrum of linear perturbations
around the standing wave (by deducing a general expression for
frequency corrections $\chi_{\lambda}$ in perturbative series
(\ref{eq:Chi_pexp_Psi_pexp})). A systematic analysis of our results
lead us to the observation that all of these corrections are given in
terms of the recurrence relation which is easy to solve.  Here we
present just a sample of our calculations for ground state solution
($\gamma=0$).  For $\chi_{0}^{+} = \omega_{\gamma=0} + \omega_{\zeta}$
the second order coefficient in (\ref{eq:Chi_pexp_Psi_pexp}) reads
\begin{widetext}
  \begin{equation}
    \chi_{2} = \frac{1134 \zeta ^6+19003 \zeta ^5+124820 \zeta ^4
      +407705 \zeta ^3+688426
      \zeta ^2+548112 \zeta +146160}{448 \zeta ^5+5600 \zeta ^4
      +25760 \zeta
      ^3+53200 \zeta ^2+47292 \zeta +13230},
  \end{equation}
  for $\zeta=0,1,\ldots$, while in the $\chi_{0}^{-} =
  \omega_{\gamma=0} - \omega_{\zeta}$ case we get
  \begin{equation}
    \begin{aligned}
      \chi_{2} = \frac{-1134 \zeta ^6-8213 \zeta ^5-16920 \zeta ^4-455
        \zeta ^3+28674 \zeta ^2+13168 \zeta -15120}{448 \zeta ^5+3360
        \zeta ^4+7840 \zeta ^3+5040 \zeta ^2-1988 \zeta -1470},
    \end{aligned}
  \end{equation}
\end{widetext}
for $\zeta=1,2,\ldots$.  Having computed also higher order terms we
can read off the asymptotic expansion of the linear spectrum of
perturbed standing wave (\ref{eq:Chi_pexp_Psi_pexp}).  Up to fourth
order in $\ep$ for large wave numbers $\zeta$ the spectrum (of ground
state standing wave solution $\gamma=0$) reads
% \begin{equation}
%   \begin{split}
%     \label{eq:Chi_plus_asymptot}
%     \Chi^{+} & = \left(2+\frac{81 \ep ^2}{32}+\frac{706663 \ep
%         ^4}{322560} + \ldots \right)\zeta \\
%     & \quad + \left(8+\frac{1207 \ep
%         ^2}{112}+\frac{908257501\ep^4}{86929920}
%       + \ldots \right) \\
%     & \quad - \left(\frac{105 \ep ^2}{64}+\frac{29319\ep ^4}{28672} +
%       \ldots \right)\zeta^{-1} \\
%     & \quad + \left( \frac{165\ep^2}{16} + \frac{472547 \ep^4}{28672}+
%       \ldots \right)\zeta^{-2} +\mathcal{O}\left(\zeta^{-3}\right)\,,
%   \end{split}
% \end{equation}
% \begin{equation}
%   \begin{split}
%     \label{eq:Chi_minus_asymptot}
%     \Chi^{-} & = -\left(2+\frac{81 \ep ^2}{32}+\frac{706663 \ep^4}{322560} + \ldots \right) \zeta \\
%     & \quad + \left(\frac{73 \ep ^2}{112}+\frac{48824929 \ep
%         ^4}{28976640}  + \ldots \right) \\
%     & \quad + \left(\frac{105 \ep ^2}{64}+\frac{29319 \ep ^4}{28672} +
%       \ldots \right)\zeta^{-1} \\
%     & \quad + \left(\frac{15\ep^2}{4} + \frac{50753\ep^4}{4096} + \ldots
%     \right)\zeta^{-2} + \mathcal{O}\left(\zeta^{-3}\right)\,.
%   \end{split}
% \end{equation}
\begin{widetext}
  \begin{equation}
    \begin{split}
      \label{eq:Chi_plus_asymptot}
      \Chi^{+} & = \left(2+\frac{81 \ep ^2}{32}+\frac{706663 \ep
          ^4}{322560} + \ldots \right)\zeta + \left(8+\frac{1207 \ep
          ^2}{112}+\frac{908257501 \ep^4}{86929920} + \ldots \right) -
      \left(\frac{105 \ep ^2}{64}+\frac{29319\ep ^4}{28672} + \ldots
      \right)\zeta^{-1} \\
      & \quad + \left( \frac{165\ep^2}{16} +
        \frac{472547\ep^4}{28672}+ \ldots \right)\zeta^{-2}
      +\mathcal{O}\left(\zeta^{-3}\right)\,,
    \end{split}
\end{equation}
\begin{equation}
  \begin{split}
    \label{eq:Chi_minus_asymptot}
    \Chi^{-} & = -\left(2+\frac{81 \ep ^2}{32}+\frac{706663
        \ep^4}{322560} + \ldots \right) \zeta + \left(\frac{73 \ep
        ^2}{112}+\frac{48824929 \ep^4}{28976640} + \ldots \right) +
    \left(\frac{105 \ep ^2}{64}+\frac{29319 \ep ^4}{28672}
      + \ldots \right)\zeta^{-1} \\
    & \quad + \left(\frac{15\ep^2}{4} + \frac{50753\ep^4}{4096} +
      \ldots \right)\zeta^{-2} + \mathcal{O}\left(\zeta^{-3}\right)\,.
  \end{split}
\end{equation}
\end{widetext}
Thus the spectrum is only asymptotically resonant for $\ep\neq
0$. This has a direct consequence on the dynamics of perturbed
standing wave solution, which we investigate in the next paragraph.

%%%%%%%%%%%%%%%%%%%%%%%%%%%%%%%%%%%%%%%%%%%%%%%%%%%%%%%%%%%%%%%%%%%%%%%%%%%%%%
\textit{Numerical results}.
%%%%%%%%%%%%%%%%%%%%%%%%%%%%%%%%%%%%%%%%%%%%%%%%%%%%%%%%%%%%%%%%%%%%%%%%%%%%%%
We solve the system (\ref{eq:wave}-\ref{eq:Phi_Pi_definitions})
subject to boundary conditions (\ref{eq:boundary}) with the same
methods as used in \cite{mr_PRL} with only minor modification due to
real and imaginary parts of dynamical fields
(\ref{eq:Phi_Pi_definitions}).  For a purely real initial data
\begin{equation}
  \begin{aligned}
    \Phi(0,x) & = 0\,, \\
    \Pi(0,x) & = \ep\frac{2}{\pi}
    \exp\left(-\frac{4\tan^{2}{x}}{\pi^{2}\sigma^{2}}\right) \,,
  \end{aligned}
\end{equation}
(with $\sigma=1/16$) we reproduce the scaling
$\mbox{$\Pi(t,0)^{2}\ra\ep^{-2}\Pi(\ep^{2}t,0)^{2}$}$ (cf. Fig.~2 in
\cite{br}) which improves with decreasing amplitude of the
perturbation $\ep$, supporting the conjectured AdS instability
\cite{br} for reflecting boundary conditions.
\begin{figure}[t]
  \centering
  \includegraphics[width=0.99\columnwidth]{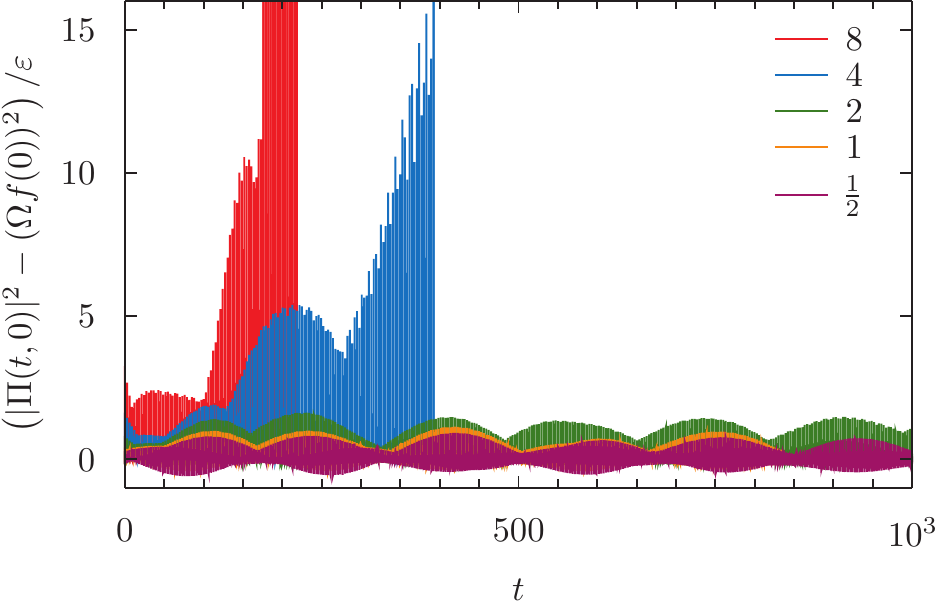}
  \\[2ex]
  \includegraphics[width=0.99\columnwidth]{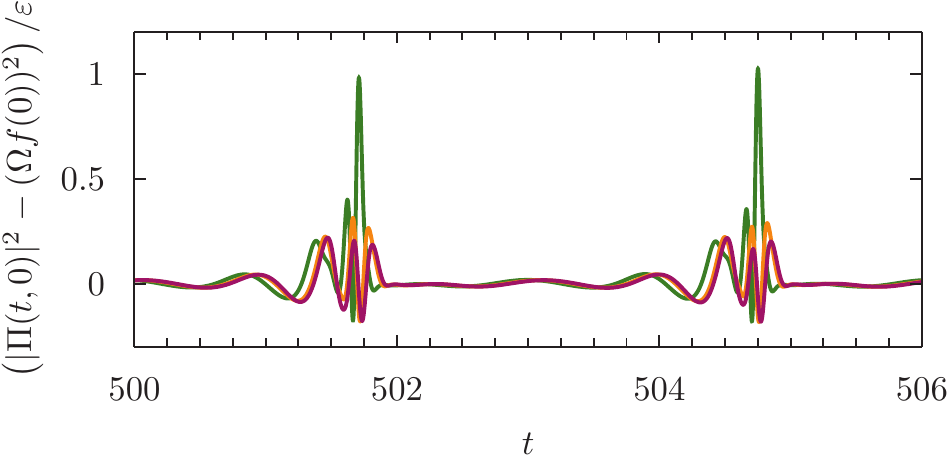}
  \caption{\textit{Top}. The time evolution (in $d=4$) of squared
    module of a scalar field $\Pi(t,x)$ at the origin ($x=0$) for a
    perturbed ground state standing wave solution with $\f(0)=0.16$
    ($\Omega\approx 4.15034$) by narrow gaussian pulse
    (\ref{eq:sw_initial_data}) of decreasing amplitude (labelled by
    different line colors). \textit{Bottom}.  A closeup showing
    scaling with an amplitude of the perturbation $\ep$, which
    improves when $\ep\ra 0$.  Because of the nonlinearity of
    governing field equations we cannot exactly separate contributions
    coming from standing wave solution and a perturbation.}
  \label{fig:ricci_sw}
\end{figure}

\begin{figure}[h]
  \centering
  \includegraphics[width=0.99\columnwidth]{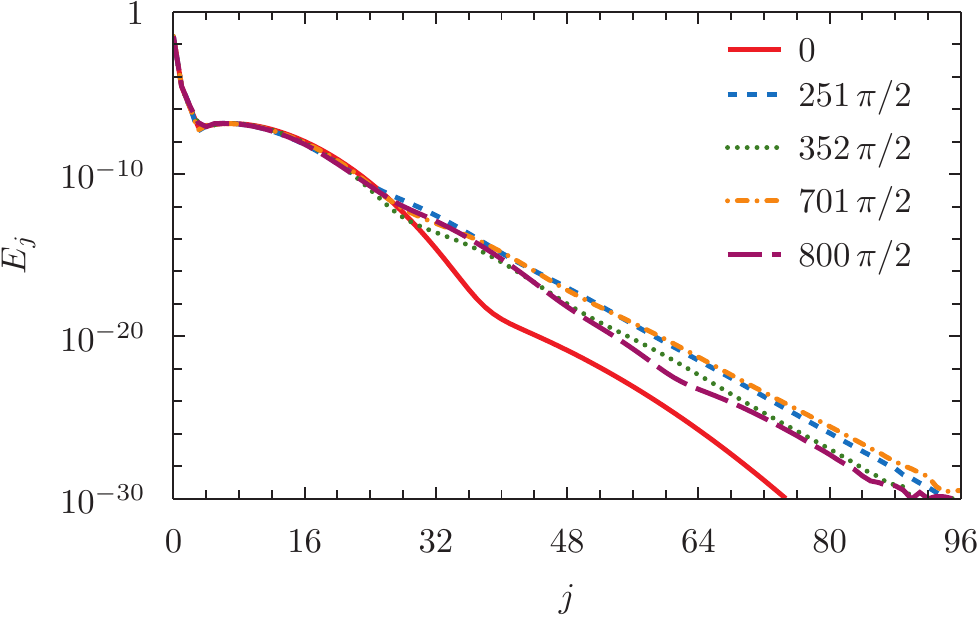}
  \caption{Plot of the energy spectrum defined as
    $\mbox{$E_{j}:=\left|\inner{e_{j}}{\sqrt{A}\Pi}\right|^{2} +
      \omega_{j}^{-2}\left|\inner{e_{j}'}{\sqrt{A}\Phi}\right|^{2}$}$
    at initial and late times (labelled by different line types) for
    the solution of perturbed standing wave (\ref{eq:sw_initial_data})
    with $\f(0)=0.16$ ($\Omega\approx 4.15034$) and amplitude of the
    gaussian perturbation $\ep=1/2$. For late times spectrum falls off
    exponentially with an almost constant slope (compare with analogue
    Fig.~2 in \cite{mr_IJMPA} for perturbed AdS solution).}
  \label{fig:spectra_sw}
\end{figure}

On the other hand for perturbed standing wave solution, i.e. for the
initial data
\begin{equation}
  \begin{aligned}
    \label{eq:sw_initial_data}
    \Phi(0,x) & = \f'(x)\,, \\
    \Pi(0,x) & = \ep\frac{2}{\pi}
    \exp\left(-\frac{4\tan^{2}{x}}{\pi^{2}\sigma^{2}}\right) +
    i\Omega\f(x)\frac{e^{\d(x)}}{\a(x)}\,,
  \end{aligned}
\end{equation}
(here we also set $\sigma=1/16$) evolution is different (see
Fig.~\ref{fig:ricci_sw} for a perturbed ground state $\gamma=0$
solution; we observe the same behaviour for small amplitude excited
states $\gamma>0$).  While for large amplitudes of the gaussian
perturbation, after several dozens of reflections, the modulus of the
scalar field $\Pi(t,0)$ starts to grow indicating formation of
apparent horizon, the situation changes when the perturbation becomes
small.  For slightly perturbed standing wave solution, and for
simulated time intervals, the evolution does not show any sign of
instability staying all the time close to the stationary
state.  Moreover, in contrast to perturbations of the pure AdS space,
in this case we do not observe any scaling with the coordinate time
$t$. Here we observe an initially narrow perturbation to bounce fourth
and back over the standing wave solution, which as a time passes tends
to spread out over the whole domain.  This effect is a consequence of
nonresonant spectra of standing wave solutions
(\ref{eq:Chi_plus_asymptot}, \ref{eq:Chi_minus_asymptot}), similarly
to the Minkowski in a cavity model with Neumann boundary
condition.  The lack of a coherence restricts the energy transfers
during successive implosions.  As a consequence, the energy spectra of
noncollapsing solution seems to equilibrate around some stationary
distribution with small fluctuation of energy between eigenmodes (see
Fig.~\ref{fig:spectra_sw}).  Therefore the solution behaves as a
perturbation propagating on a standing wave background as is seen on
Fig.~\ref{fig:ricci_sw} where we subtract a contribution of standing
wave (a constant value) and rescale by the amplitude of initial
perturbation.

%%%%%%%%%%%%%%%%%%%%%%%%%%%%%%%%%%%%%%%%%%%%%%%%%%%%%%%%%%%%%%%%%%%%%%%%%%%%%%
\textit{Conclusions}.
%%%%%%%%%%%%%%%%%%%%%%%%%%%%%%%%%%%%%%%%%%%%%%%%%%%%%%%%%%%%%%%%%%%%%%%%%%%%%%
There is growing evidence that while the AdS space is unstable
against a black hole formation under a large class of arbitrarily
small initial perturbations \cite{br, jrb, dhs, bll}, there also
exists a variety of stable, asymptotically AdS (aAdS) solutions like
time-periodic solutions \cite{mr_PRL} or standing waves \cite{bll_BS},
that can be arbitrarily close to AdS. Both AdS and those aAdS
solutions are stable at linear level.
% (here we have considered only stable branch of standing waves,
% i.e. solutions with small $\f(0)$, on the left of first local
% extreme of total energy function; we observe similar behaviour as
% for an asymptotically flat boson stars \cite{hc}; detailed analysis
% of moderate and large $\f(0)$ will be discussed elsewhere
% \cite{m_sw}).
However, a small perturbations in a form of a short pulse of
radiation, when perturbing the pure AdS solution, propagates roughly
non-dispersively, with its energy cascading to higher frequencies and
ultimately collapsing to a black hole, while perturbing those aAdS
solutions (e.g. time-periodic ones), it disperses over the whole space
and the energy cascade is ultimately stopped. It was suggested in
\cite{dhms} that this qualitative change in the long time evolution is
due to the nonresonant character of the spectrum of linear
perturbations of the aAdS solutions. In this note, as the first step
to study the stability of time-periodic solutions, we investigate in
detail the spectrum of linear perturbations of standing waves
\cite{bll_BS} and show that indeed it is not exactly resonant (it is
only asymptotically resonant --- resonant in the limit of a wave
number going to infinity). Studying this system numerically we find
that there is a threshold for triggering instability resulting in a
black hole formation. It is important to stress that such stable aAdS
solutions can be arbitrarily close to AdS. Then the fate of a small
perturbations in a form of a short pulse of radiation depends on what
dominates as perturbation of pure AdS: if ``short pulse'' dominates
over a standing wave or a time-periodic solution (there is a ``short
pulse'' perturbed with some small stable aAdS solution) it will still
trigger the energy cascade and ultimately collapse to a black hole; if
some stable aAdS solution dominates over a ``short pulse'' (there is a
stable aAdS solution perturbed with a ``short pulse'') the energy
cascade is stopped and the evolution stays smooth. Then we confirm
this qualitative behavior in the toy model of a portion of Minkowski
space $\Lambda=0$ enclosed in a perfectly reflecting cavity. The
advantage of this somehow artificial model is that it allows for the
two types of boundary conditions, resulting in either resonant or only
asymptotically resonant spectrum of linear perturbations (for
Dirichlet and Neumann boundary conditions respectively). Indeed the
long time evolution of small perturbations for these two types of
boundary conditions is qualitatively different. For the Dirichlet
boundary conditions the perturbation ultimately collapses to a black
hole (an analogue of the pure AdS case), while for Neumann boundary
conditions (an analogue of a stable aAdS solution) there is a
threshold for a black hole formation --- the small perturbations do
not collapse and their evolution stays smooth.

%%%%%%%%%%%%%%%%%%%%%%%%%%%%%%%%%%%%%%%%%%%%%%%%%%%%%%%%%%%%%%%%%%%%%%%%%%%%%%
\vskip 0.1cm \noindent \emph{Acknowledgments:}
%%%%%%%%%%%%%%%%%%%%%%%%%%%%%%%%%%%%%%%%%%%%%%%%%%%%%%%%%%%%%%%%%%%%%%%%%%%%%%
We are indebted to Piotr Bizo\'n for suggestions and discussions. This
work was supported by the NCN grant
DEC-2012/06/A/ST2/00397. M.M. acknowledges support from the Dean's
grant no.~K/DSC/001588.  The computations were performed at the
supercomputer ``Deszno'' purchased thanks to the financial support of
the European Regional Development Fund in the framework of the Polish
Innovation Economy Operational Program (contract
no. POIG.~02.01.00-12-023/08) and at the ``Mars'' supercomputer of
Academic Computer Centre Cyfronet AGH, grant
no. MNiSW/IBM\_BC\_HS21/UJ/071/2013.

\end{document}